\documentclass[aps,prl,showpacs,twocolumn,groupedaddress]{revtex4}
\usepackage{graphicx}
\begin{document}
%\preprint{}


\title{Measurement of electron correlations in Li$_{x}$CoO$_{2}$ ($x=$0.0 $-$ 0.35) using $^{59}$Co nuclear magnetic resonance and nuclear quadrupole resonance techniques}

\author{S. Kawasaki$^1$}
%\email[]{kawasaki@science.okayama-u.ac.jp}
\author{T. Motohashi$^{2,3}$}
\author{K. Shimada$^1$}
\author{T. Ono$^{2,4}$}
\author{R. Kanno$^4$}
\author{M. Karppinen$^{2,5}$}
\author{H. Yamauchi$^{2,4,5}$}
\author{Guo-qing Zheng$^1$}

\affiliation{$^1$Department of Physics, Okayama University, Okayama 700-8530, Japan}
\affiliation{$^2$Materials and Structures Laboratory, Tokyo Institute of Technology, Yokohama 226-8503, Japan}
\affiliation{$^3$Graduate School of Engineering, Hokkaido University N13, W8, Kita-ku, Sapporo 060-8628, Japan}
\affiliation{$^4$Interdisciplinary Graduate School of Science and Engineering, Tokyo Institute of Technology, Yokohama 226-8502, Japan}
\affiliation{$^5$Laboratory of Inorganic Chemistry, Department of Chemistry, Helsinki University of Technology, FI-02015 TKK, Finland}


%\author{}
%\thanks{}
%\affiliation{}


%\date{\today}

\begin{abstract}
CoO$_{2}$ is the parent compound for the superconductor Na$_{x}$CoO$_{2}\cdot$1.3H$_{2}$O and  was widely believed to be a Mott insulator.  
We performed $^{59}$Co nuclear magnetic resonance (NMR) and nuclear quadrupole resonance (NQR) studies on Li$_x$CoO$_2$ (x = 0.35, 0.25, 0.12, and 0.0) to uncover the electronic state and spin correlations in this series of compounds which was recently obtained through electrochemical de-intercalation of Li from pristine LiCoO$_2$. We find that although the antiferromagnetic spin correlations systematically increase with decreasing Li-content ($x$), the end member, CoO$_{2}$ is  a non-correlated metal that well satisfies the Korringa relation for a Fermi liquid. Thus, CoO$_2$ is not simply located at the limit of $x\to$0 for A$_x$CoO$_2$ (A = Li, Na) compounds.   The disappearance of the electron correlations in CoO$_2$ is due to the three dimensionality of the compound which is in contrast to the highly two dimensional structure of A$_x$CoO$_2$.

\end{abstract}

\pacs{74.25.Jb, 74.70.-b, 74.25.Nf}
%\keywords{}

\maketitle


%************************intruduction*************************************************************************************
Electronic correlations and superconductivity in transition-metal oxides have been a main focus in condensed matter physics since the discovery of high transition-temperature ($T_{\rm c}$) superconductivity in copper oxides. 
The hydrated cobalt-oxide  superconductor Na$_{x}$CoO$_{2}\cdot$1.3H$_{2}$O has intensified the research interest in the past few years\cite{Takada}.
This compound bears similarities to the high-$T_{\rm c}$ copper oxides in that it has a quasi-two dimensional crystal structure and contains a transition-metal element that carries a spin of $\frac{1}{2}$. 
Indeed, nuclear quadrupole resonance (NQR)  measurements on Na$_{x}$CoO$_{2}\cdot$1.3H$_{2}$O have found $T^3$ variation below $T_{\rm c}$ in the spin-lattice relaxation rate $1/T_1$, which is a strong indication of existence of line nodes in the superconducting gap function \cite{Fujimoto,Zheng,Kusano}. 
Precise measurements of the Knight shift in a high quality single crystal reveals that the spin susceptibility decreases below $T_{\rm c}$ along both $a$- and $c$-axis directions, which indicates that the Cooper pairs are in the spin-singlet state \cite{Zheng2}. Thus, the superconductivity in Na$_{x}$CoO$_{2}\cdot$1.3H$_{2}$O appears to be of $d$-wave symmetry, as in the case of high-$T_{\rm c}$ copper oxides. It has also been found that antiferromagnetic spin correlations are present in the superconducting cobaltates, though being much weaker than those in the cuprates \cite {Fujimoto,Zheng}. The correlations are anisotropic in the spin space \cite{Matano}, which is different from the cuprate case.

Then, a natural question is how to model the cobalt oxides. Many authors applied the so-called $t-J$ model that had been widely used to describe the cuprates \cite{Baskaran,WangLeeLee,Ogata}. In these theories, one virtually starts from CoO$_2$ in which Co is in the Co$^{4+}$ state and there is one electron ($s$ = 1/2) in the lowest level ($a_{1g}$ orbital). Upon adding Na, one dopes electrons into the $a_{1g}$ orbital, and creates a Co$^{3+}$ ($s$ = 0)  state.  In such a scenario, one may be in a situation of dealing with a doped Mott insulator, as in the cuprates case \cite{Baskaran,WangLeeLee,Pickett}. Therefore, it is important to synthesize the CoO$_2$ phase and reveal its electronic ground state. Unfortunately, it has been chemically difficult to obtain pure phase of CoO$_2$, or even Na$_x$CoO$_2$ with $x\leq$ 0.25, though some efforts have been reported \cite{Julien,Sugiyama,Tarascon}.

In this paper, we report $^{59}$Co NMR and NQR studies to uncover the electronic state and spin correlations in Li-deficient phases, Li$_{x}$CoO$_{2}$ ($x$ = 0.35, 0.25, and 0.12), and the CoO$_2$ phase. Although the antiferromagnetic spin correlation increases with reducing Li-content ($x$), the end member, CoO$_{2}$ is found to be a non-correlated metal that well satisfies the Korringa relation for a Fermi liquid. The result obtained from our CoO$_2$ sample is different from the one reported earlier \cite{Julien} in both the temperature ($T$) dependence and the magnitude of the $1/T_1$. It turns out that the earlier result correspond to that of our Li$_{0.12}$CoO$_{2}$.
We argue that, however, the disappearance of the electron correlations in pure CoO$_{2}$ is due to the three dimensionality of the compound which collapses from the highly two dimensional structure of A$_x$CoO$_2$ (A = Li, Na) when Li is completely removed. The systematic evolutions of the electron correlations in Li$_{x}$CoO$_{2}$ ($x$ = 0.35, 0.25, and 0.12), as well as in Na$_{x}$CoO$_{2}\cdot$ 1.3H$_{2}$O ($x$ = 0.35, 0.33, 0.28, and 0.25) \cite{Zheng}, are consistent with the theoretical postulation that A$_x$CoO$_2$ (A = Li, Na) with small $x$ be near a magnetic instability \cite{Baskaran,WangLeeLee}.

%************************experiment*********************************************
\begin{figure}[h]
\includegraphics[width=6cm]{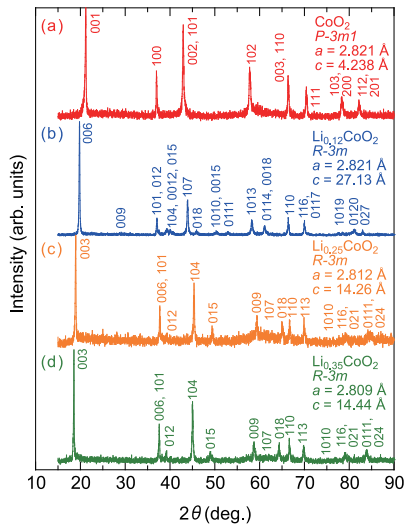}% Here is how to import EPS art
\caption{ (Color online) X-ray powder diffraction patterns for (a) CoO$_2$, (b) Li$_{0.12}$CoO$_2$, (c) Li$_{0.25}$CoO$_2$, and (d) Li$_{0.35}$CoO$_2$ samples. For these samples, Rietveld refinement of the crystal structure was unsuccessful, as the use of an airtight sample holder had significantly deteriorated the resolution of the diffraction patterns. }
\end{figure}

Polycrystalline samples of Li$_x$CoO$_2$ ($x$ = 0.35, 0.25, and 0.12) and CoO$_2$ ($x$ = 0.0) were synthesized through electrochemical de-intercalation of Li from pristine LiCoO$_2$, as described elsewhere \cite{Motohashi,Motohashi2}. Approximately 100 mg of single-phase LiCoO$_2$ pellet (without additives) was electrochemically oxidized with a constant current of 0.1 mA in an airtight flat cell filled with a nonaqueous electrolyte. The Li content (or the amount of Li ions to be extracted, i.e. 1-$x$) of each sample was precisely controlled by the reaction duration based on Faraday's law. Typically, a 100-mg LiCoO$_2$ pellet was charged for 178, 205, 241, and 274 h to obtain the $x$ = 0.35, 0.25, 0.12, and 0.0 (i.e. CoO$_2$) phases, respectively. As seen in Fig.1, x-ray powder diffraction analysis evidenced that all the samples are of single phase with characteristic crystal structures typical for their Li compositions. Sharp diffraction peaks throughout the XRD patterns demonstrate that our Li$_x$CoO$_2$ and CoO$_2$ samples are chemically homogenous with good crystallinity. The actual $x$ values determined by inductively coupled plasma-atomic emission spectroscopy (ICP-AES) were in excellent agreement with the theoretical ones, indicating that the full amount of electricity due to the current was used for Li de-intercalation from LiCoO$_2$. Since high-valent cobalt oxides tend to experience chemical instability when exposed to atmospheric moisture, sample handling and characterization were carefully made in an inert gas atmosphere. A part of the electrochemically-treated samples ($\sim$70 mg) was encapsulated into a Pyrex ampule filled with Ar gas. NMR/NQR measurements were performed by using a phase coherent spectrometer. The NQR measurements were performed at zero magnetic field. The NMR and NQR spectra were taken by changing the external magnetic field ($H$) at a fixed rf frequency of 71.1 MHz and by changing rf frequency and recording the spin echo intensity step by step, respectively.  The value of $1/T_1$ was extracted by fitting the nuclear magnetization obtained by recording the spin echo intensity to the Master equation\cite{Maclaughlin,Chepin}.

%************************spectra and K******************************************%\begin{figure}[h]

\begin{figure}[h]
\includegraphics[width=8.5cm]{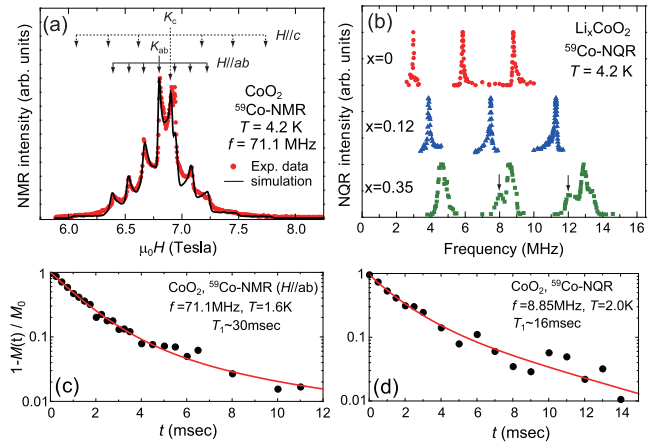}% Here is how to import EPS art
\caption{ (Color online) (a) $^{59}$Co NMR spectra for CoO$_2$ measured at $T$ = 4.2 K. The NMR frequency is 71.1 MHz. Solid and dotted arrows indicate the two sets of seven NMR peaks originated from anisotropy of the Knight shift, respectively.  (b) NQR spectrum for Li$_{x}$CoO$_{2}$ ($x$ = 0.0, 0.12, and 0.35) measured at $T$ = 4.2 K. Arrows indicate extrinsic NQR peaks occurred due to aging (degradation). (c) and (d) are typical data sets of $^{59}$Co nuclear recovery curves for CoO$_2$ obtained by NMR and NQR, respectively (see text).}
\end{figure}

Figure 2 (a) shows a representative $^{59}$Co-NMR spectrum for CoO$_{2}$. The spectrum shows a typical randomly-oriented powder pattern. Since $^{59}$Co nucleus has  nuclear spin $I$ = 7/2, an NMR spectrum has seven peaks due to the quadrupole interaction\cite{Abragam}. As schematically shown by solid and dotted arrows in Fig.2(a), the observed spectrum consists of  two central peaks originated from the anisotropy of the Knight shift along the $ab$- ($K_{\rm ab}$) and $c$- ($K_{\rm c}$) directions and each central peak is accompanied by six satellite peaks.  The numerical calculation considering the quadrupole effect up to the second-order perturbation completely reproduced the experimental result (solid curve in Fig.2 (a)). The clear peak structure attests the high quality of the sample.   Thus, we are able to obtain the values of $K_{\rm ab}$ and $K_{\rm c}$ precisely from the NMR spectra.  Figure.2 (b) shows the $^{59}$Co-NQR spectra for Li$_x$CoO$_2$ ($x$ = 0.0, 0,12, and 0.35) observed at zero magnetic field. As in  Na$_x$CoO$_2$$\cdot$yH$_2$O\cite{Fujimoto,Zheng}, three NQR transition lines arising from  $I$ = 7/2  are clearly observed in Li$_x$CoO$_2$. For $x$ = 0.35, satellite peaks are observed as indicated by the arrows in Fig.2(b). It indicates that a secondary  phase is present in this composition although X-ray diffraction immediately after sample synthesis showed a single-phase pattern. Since these peaks increase in intensity as time elapsed (not shown), this is an extrinsic phase that arises after the X-ray diffraction analysis.

  Figures 2 (c) and (d) show typical datasets of $^{59}$Co nuclear recovery curves to obtain $T_1$ by NMR and NQR, respectively. As drawn in solid curves in figures, they can be fitted by single component of theoretical curves\cite{Maclaughlin,Chepin}, even though the $T_1$ is measured in powdered sample. Compared to the early report in which the NMR spectrum  did not show clear peak structure since it was a superposition of signals from different phases and $T_1$ is not of single component \cite{Julien}, it is obvious that the present sample has much better quality, and the result represents, we believe, the intrinsic  property of CoO$_2$.

The NQR parameters are summarized in Table.1. Here $\nu_{\rm Q}$ and asymmetry parameter $\eta$ are defined as $\nu_{\rm Q}$ $\equiv$ $\nu_{\rm z}$ = $\frac{3}{2I(2I-1)h}$$e^2Q$$\frac{\partial ^2V}{\partial z^2}$, $\eta$ = $\frac{|\nu_x - \nu_y|}{\nu_z}$, with $Q$ and $\frac{\partial ^2V}{\partial \alpha^2}$ ($\alpha$ = $x$, $y$, $z$) being the nuclear quadrupole moment and the electric field gradient (EFG) at the position of the Co nucleus, respectively.\cite{Abragam} Notably, $\nu_{\rm Q}$  increases with increasing $x$. This assures  electron doping by the increasing of Li-content. On the other hand, $\eta$ is almost the same in $x$ = 0.12 and 0.35, but is substantially reduced in CoO$_2$. This is because the CoO$_2$ phase crystallizes in a simple structure containing CoO$_2$ layers only (the so-called O1-type structure), while the crystal of Li$_x$CoO$_2$ consists of alternate stacking of Li$_x$ and CoO$_2$ blocks.  \cite{Motohashi}.

% •\'Ì'}"ü
\begin{table}[h]
 \caption{NQR parameters for Li$_x$CoO$_2$ obtained at 4.2 K.}% {}"à'É•\'è'ð''­
 \begin{center}
  \begin{tabular}{ccc}
    \hline
    \hline
        sample            & $^{59}$$\nu_{\rm Q}$ (MHz)   & $\eta$   \\
    \hline
      CoO$_2$             & 2.93   &  0.05$\pm$0.01  \\
    \hline
     Li$_{0.12}$CoO$_2$   &  3.76  &  0.09$\pm$0.01  \\
         
     Li$_{0.35}$CoO$_2$  &  4.32  &  0.10$\pm$0.02  \\
    \hline
    \hline  
  \end{tabular}
 \end{center}
\end{table}

\begin{figure}[h]
\includegraphics[width=6cm]{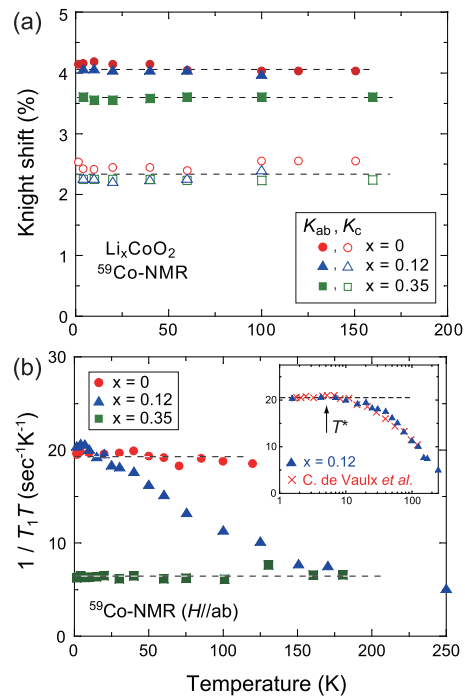}% Here is how to import EPS art
\caption{\label{fig:shift} (Color online)
(a) $T$ dependence of the Knight shift ($K$) for Li$_x$CoO$_2$ ($x$ = 0.35, 0.12, and 0.0) along $ab$-(solid symbols) and $c$-axis (open symbols), respectively. Dotted lines indicate the relation of $K$ = constant. (b) $T$ dependence of $^{59}$Co-NMR 1/$T_1T$ for Li$_x$CoO$_2$ ($x$ = 0.35, 0.12, and 0.0) measured at the field along the $a(b)$-axis. Inset shows the semi-log plot of $T$ dependence of $^{59}$Co-NMR 1/$T_1T$ for Li$_{0.12}$CoO$_2$ ($H$$\parallel$$ab$) along with the data referred from Ref. \cite{Julien}. Dotted lines indicate the relation of $1/T_1T$ = constant. Arrow indicates $T^*$. (see text)    }
\end{figure} 
Figures 3 (a) shows the $T$ dependence of the Knight shift ($K_{\rm ab}$ and $K_{\rm c}$) for three samples with different Li content. As clearly seen in the figure, both $K_{\rm ab}$ and $K_{\rm c}$ of Li$_x$CoO$_2$ do not depend on temperature.  
Here, the Knight shift consists of contributions from the spin susceptibility, $K_{\rm s}$, and from the orbital susceptibility (Van Vleck susceptibility), $K_{\rm orb}$.
$K = K_s + K_{orb}$,
with $K_{\rm orb}$ being $T$-independent but $K_{\rm s}$ being $T$-dependent generally.  $K_{\rm s}(T)$ and $K_{\rm orb}$ are respectively related to the spin susceptibility $\chi_{\rm s}$ and orbital susceptibility $\chi_{\rm orb}$ as $K_s(T) = A_{hf} \chi_s(T)$ and $K_{orb} = A_{orb} \chi_{orb}$, where $A_{\rm hf}$ is the hyperfine coupling constant between the nuclear and the electron spins. The results show that the spin susceptibility in Li$_x$CoO$_2$ is $T$-independent. 

%This is consistent with the results of previous bulk susceptibility measurements which conclude that the ground state of CoO$_2$ is a Pauli paramagnet.\cite{Motohashi,Motohashi2} 

Figure 3 (b) shows the $T$ dependence of  $1/T_1T$ measured by $^{59}$Co-NMR with $H$$\parallel$$ab$. Surprisingly, the $1/T_1T$ for  CoO$_2$ is $T$-independent. Together with the $T$-independent Knight shift in CoO$_2$, the Korringa relation is satisfied as discussed later in more detail. This is a strong and the first evidence for a weakly-correlated ground state of CoO$_2$.   

De Vaulx $et$ $al$ \cite{Julien} suggested that CoO$_2$ is a strongly correlated system on the basis of a small value of the characteristic temperature, $T^*$, below which the Korringa relation holds. However, as seen in Fig.3 (b) inset, we find that their result is almost the same as that for  our Li$_{0.12}$CoO$_2$ sample. As the authors acknowledged \cite{Julien}, their sample contained a Li-rich phase as impurity. We suggest that the present results clarify, for the first time, the true electronic state of CoO$_2$.

\begin{figure}[h]
\includegraphics[width=7cm]{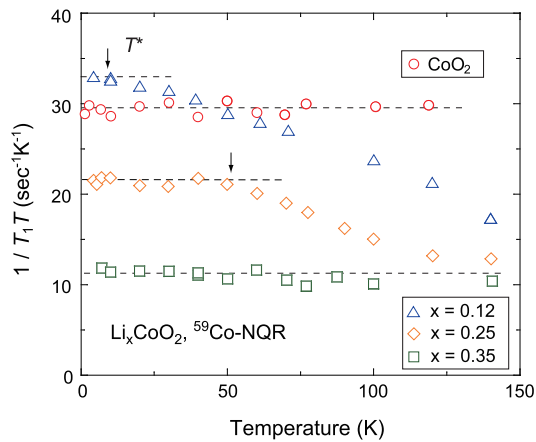}% Here is how to import EPS art
\caption{\label{fig:t1t} (Color online) $T$ dependence of $^{59}$Co-NQR 1/$T_1T$ for Li$_{x}$CoO$_{2}$ ($x$ = 0.35, 0.25, 0.12, and 0.0) Dotted lines and arrows indicate the relation of $1/T_1T$ = constant and $T^*$, respectively.  }
\end{figure}
%

%************************NQRT1T***************************************************************************************************
We further measured the  $T_1$ systematically by $^{59}$Co-NQR at zero magnetic field, which corresponds to the configuration of $H\parallel c$-axis since the principal axis of the EFG is along the $c$-axis.        
As shown in Fig.4, the $T$ dependence of $1/T_1T$ shows systematic change with decreasing Li-content, except for CoO$_2$ which will be discussed separately later. With decreasing $x$ from 0.35 to 0.12, $1/T_1T$ increases with decreasing $T$, indicating that the electron correlations are induced. Also, the increasing is more pronounced in samples with smaller $x$, which indicates that the spin correlation is stronger in samples with smaller $x$. A similar situation was encountered in Na$_{x}$CoO$_{2}\cdot$1.3H$_{2}$O ($x$ = 0.35, 0.33, 0.28, and 0.25) \cite{Zheng}. As in that case, the correlation is antiferromagnetic in origin since the Knight shift is $T$-independent. The $1/T_1T$ becomes constant below $T^*$, indicating a renormalized Fermi liquid state below $T^*$. 
This situation resembles that in electron-doped cuprate Pr$_{0.91}$LaCe$_{0.09}$CuO$_4$ where $T^* \sim$60 K \cite{Zhengelec}.
 Furthermore, $T^*$ decreases  from 50 K for $x$ = 0.25 to 7 K for $x$ = 0.12. This also indicates that a sample with smaller $x$ is closer to a magnetic  instability.  Therefore, the results are consistent with the theories for a compound near a magnetic transition \cite{Baskaran,WangLeeLee}.
% 
%Such a situation is very similar to heavy-fermion systems in which pressure is the tuning parameter instead of doping \cite{Shinji}.

However, $1/T_1T$ is  constant  for CoO$_2$. This abrupt change in the electronic state is clearly  due to the abrupt change in the crystal structure.  
The Li$_x$CoO$_2$ phase with finite $x$ has a highly two-dimensional crystal structure in which the interlayer Co-Co distance ($d_{Co-Co}$) is as large as 5.0-5.1 $\AA$, while CoO$_2$ crystallizes in a less anisotropic structure. Since there is no "spacer" layer between two adjacent CoO$_2$ blocks when Li ions are completely extracted, the $d_{Co-Co}$ value is reduced to 4.24 $\AA$ in CoO$_2$. 
The emergent three dimensionality is believed to be the origin of the weak electron correlation of CoO$_2$.

Finally, we examine if there exists any renormalization effect in CoO$_2$. To this end, we evaluate the Korringa ratio,
$S=T_1TK_s^2 \cdot\frac{4\pi k_B}{\hbar}(\frac{\gamma_n}{\gamma_e})^2$.
This quantity is  unity for a free electron system. It is much smaller than the unity for a antiferromagnetically correlated metal but much larger than the unity for a ferromagnetically correlated metal \cite{Moriya}.
In the present case, we use the $K_{\rm orb}^a$ = 2.96\% and $K_{\rm orb}^c$ = 1.72\% obtained from recent NMR study in single crystalline Na$_{0.42}$CoO$_2$ \cite{Matano}, then we obtain $S$ = 1.12$\pm$0.04 for CoO$_2$. 
Therefore, {\it CoO$_2$ is a conventional metal that well conforms to  Fermi liquid theory}.

%************************summary*************************************************************************************
In conclusion, we have presented $^{59}$Co-NMR and NQR measurements and analysis on  Li$_{x}$CoO$_{2}$ ($x=$0.0$-$0.35). 
The antiferromagnetic-like spin fluctuations develop when Li is de-intercalated from Li$_{0.35}$CoO$_{2}$, which is consistent with the picture that the member of the families A$_x$CoO$_2$ (A = Li, Na) with small $x$ be viewed as a doped spin 1/2 system.    Due to the emergent three-dimensionality of the crystal structure, however, CoO$_2$, the $x$ = 0 end member of A$_x$CoO$_2$, is a conventional metal that well conforms to  Fermi liquid theory. The result highlights the importance of two dimensionality for electron correlations in A$_x$CoO$_2$, as was the case that water intercalated into non-correlated Na$_{0.42}$CoO$_2$ brings about spin fluctuations \cite{Matano}. We hope that these results form a foundation for understanding   cobalt oxides and the  superconductivity developed out of there.

%************************acknowledgments*************************************************************************************
This work was supported  in  part  by  research grants from  MEXT and JSPS, and also by Tekes (No. 1726/31/07) and Academy of Finland (No. 110433).

%\bibliography{apssamp}

\end{document}